\documentstyle[12pt]{article}
\newlength{\extraspace}
\newlength{\extraspaces}
\newcommand{\be}{\begin{equation}
\addtolength{\abovedisplayskip}{\extraspaces}
\addtolength{\belowdisplayskip}{\extraspaces}
\addtolength{\abovedisplayshortskip}{\extraspace}
\addtolength{\belowdisplayshortskip}{\extraspace}}
\newcommand{\ee}{\end{equation}}

\newcommand{\ba}{\begin{eqnarray}
\addtolength{\abovedisplayskip}{\extraspaces}
\addtolength{\belowdisplayskip}{\extraspaces}
\addtolength{\abovedisplayshortskip}{\extraspace}
\addtolength{\belowdisplayshortskip}{\extraspace}}
\newcommand{\ea}{\end{eqnarray}}

\newcommand{\nonu}{\nonumber \\[.5mm]}
\newcommand{\A}{&\!\!\!}

\def\thesection {\S {\arabic{section}}}
\newcommand{\newsection}[1]{
\vspace{7mm}
\pagebreak[3]
\addtocounter{section}{1}
\setcounter{subsection}{0}
\setcounter{footnote}{0}
\begin{center}
{\large {\bf \thesection. #1}}
\end{center}
\nopagebreak
\medskip
\nopagebreak
\hspace{3mm}}

\begin{document}

%%%%%%%%%%%%%%%%%%%%%%%%%%%%%%%%%%%%%%%%%%%%%%%%%%%%%%%%%%%%%%%%%%%%%%%%%
%%%%%%%%%%%%%%%%%%%%%%%%%%%%%%%%%%%%%%%%%%%%%%%%%%%%%%%%%%%%%%%%%%%%%%%%%
%%%%%%%%%%%%%%%%%%%%%%%%%%%%%%%%%%%%%%%%%%%%%%%%%%%%%%%%%%%%%%%%%%%%%%%%%
\newpage

\begin{large}

\centerline{\bf Energy and Momentum in the Tetrad Theory of Gravitation}
%\centerline{\bf the Tetrad Theory of Gravitation}

\end{large}

\hspace{2cm}

%\begin{small}

\centerline{Takeshi S{\footnotesize HIRAFUJI} and Gamal G. L. N{\footnotesize ASHED}}
\centerline{{\it Physics Department, Saitama University, Urawa 338}}

\bigskip

%\end{small} 

\thispagestyle{empty}

\hspace{2cm}
\\
\\
\\
\\
\\
\\
\\
\\

We study the energy and momentum of an isolated system in the tetrad theory of gravitation, 
starting from the most general Lagrangian  quadratic in torsion,  which involves four unknown 
parameters. When  applied to the static spherically symmetric case, the parallel vector fields take
 a diagonal form, and the field equation has an exact solution. We analyze the linearized field 
equation in vacuum at distances far from the isolated system without assuming any symmetry
 property of the system. The linearized equation is a set of coupled equations for a symmetric 
and skew-symmetric tensor fields, but it is possible to solve it up to $O(1/r)$ for the stationary 
case. It is found that the general solution contains two constants, one being the gravitational 
mass of the source and the other  a constant vector ${\grave B_\alpha}$. The total energy is 
calculated from this solution and is found to be equal to the gravitational mass of the source. 
We also calculate the spatial momentum  and  find that its value  coincides with the constant 
vector ${\grave B_\alpha}$. The linearized field equation in vacuum, which is valid at  
distances far from the source, does not give any information about whether the constant vector 
${\grave B_\alpha}$ is vanishing or not. For a weakly gravitating source for which the field is
 weak everywhere, we find that the constant vector ${\grave B_\alpha}$ vanishes.
%%%%%%%%%%%%%%%%%%%%%%%%%%%%%%%%%%%%%%%%%%%%%%%%%%%%%%%%%%%%%%%%%%%%%%%
%\setcounter{page}{1}
%\setcounter{section}{0}
%\setcounter{equation}{0}

\newpage
\newsection{Introduction}

Einstein concerned himself with the problem of energy and momentum for a system of matter 
plus gravitational field shortly after he proposed general relativity.$^{1)}$ He introduced the 
well-known expression for the energy-momentum complex ${\theta_{\mu}}^{\nu}$, which is referred 
to as a 
canonical expression and satisfies ordinary conservation law as a consequence of the gravitational 
field equation. For a closed system
and for a restricted class of coordinates the quantities $P_\mu$ obtained from the complex 
${\theta_{\mu}}^{\nu}$ by integrating over spatial coordinates satisfy the following properties:$^{2), 3)}$ (1) They are constant in time, (2) transform as a 4-vector under  linear coordinate 
transformations, and (3) are invariant under arbitrary transformations of spatial coordinates 
which tend to the identity transformation at infinity. Furthermore (4) the gravitational mass 
of the system is equal to the total energy. It has recently been shown that a characteristic 
property of energy thus defined is its positivity, reflecting stability of the physical 
${\rm system.}^{4)-6)}$ 

 M\o ller revived$^{7)}$ the issue of energy and momentum in general relativity, and required$^{8)}$ that any energy-momentum complex 
 ${\tau_{\mu}}^{\nu}$ must satisfy the following properties: 
(A) It must be an affine tensor density which satisfies conservation law, (B) for an isolated  system the quantities 
 $P_{\mu}$ are constant in time and transform as the covariant components of a 4-vector under linear coordinate transformations, 
  and  (C) the superpotential  ${{\cal U_{\mu}}^{\nu \lambda}}$ $=-
  {{\cal U_{\mu}}^{\lambda \nu}}$ transforms as a tensor density of rank 3 under the group of spacetime transformations. The second property requires that ${\tau_\mu}^{\nu}$ around an isolated  system should decrease faster than $1/r^3$ for large values of the radial coordinate $r$. The last property is required in order that the four-momentum for any isolated system be a 4-vector under arbitrary spacetime transformations.
 
 The canonical energy-momentum complex ${\theta_\mu}^{\nu}$ does not satisfy (C). An 
energy-momentum complex that satisfies properties (A) and (C) was ${\rm constructed,}^{7),9)}$ but 
it was found later$^{10)}$ that the property (B) is not satisfied. 
${\rm Lessner}^{11)}$ discussed that although this result is inadequate from the viewpoint of
 special relativity, it may be adequate from the viewpoint of general relativity. 
 
It is not possible to satisfy all the above requirements if the gravitational field is 
described by the metric tensor alone.$^{10)}$ In a series of 
${\rm papers,}^{10),12),13)}$ therefore,  M\o ller was led to the tetrad 
description of 
gravitation, and constructed a formal form of energy-momentum complex that satisfies all the requirements. The metric tensor is uniquely fixed by the tetrad field, but the reverse is not true, since the tetrad has six extra degrees of freedom. In the tetrad formulation of general relativity, the tetrad field is allowed to undergo local Lorentz transformations with six arbitrary functions. The energy-momentum complex is not a tensor and changes its form under such  transformations. Therefore, unless one can find a good physical argument for fixing the tetrad throughout the system, one cannot speak about the energy distribution inside the system. The total energy-momentum obtained by the complex, however, is invariant under local Lorentz transformations with appropriate boundary conditions.$^{13)}$

M\o ller also suggested another possibility that the tetrad field is uniquely fixed by means
 of six supplementary conditions.$^{12)}$ In this case the underlying spacetime possesses 
absolute parallelism,$^{14)}$ with the tetrad field playing the role of  the parallel vector 
fields, which are allowed to undergo only global  Lorentz transformations. The Lagrangian 
formulation of this tetrad theory of gravitation  was first given by Pellegrini 
and Plebanski.$^{15)}$  Hayashi and ${\rm Nakano}^{16)}$ independently formulated the same 
gravitational theory as a gauge theory of spacetime translation group.   
The Lagrangian was assumed to be given by a sum of quadratic invariants of the torsion tensor, 
which is expressed by first-order derivatives of the parallel vector fields. Its most general 
expression then involves four unknown parameters to be determined by experiment, which we denote
 here as $a_1$, $a_2$, $a_3$ and $a_4$. The last parameter $a_4$ is associated with a
 parity-violating term. At first it was required that for the weak field case the gravitational 
field equation should reproduce the linearized Einstein equation. This restricted the parameters 
as $a_1 + a_2 = 0$ and $a_4 = 0$.$^{16),17)}$    
M\o ller$^{17)}$ also suggested a possible generalization of the gravitational Lagrangian by 
including homogeneous functions of the torsion tensor of degree 4 or  higher. 

Hayashi and ${\rm Shirafuji}^{18)}$ studied the geometrical and observational basis of the 
tetrad theory of gravitation\footnote {They coined the name "new general relativity", because 
Einstein$^{19)}$ was the first to introduce the notion of absolute parallelism into physics 
after he had constructed general relativity.}$^)$ assuming  the Lagrangian to be invariant under 
parity operation, involving three unknown parameters $a_1$, $a_2$ and $a_3$. Two of these 
parameters $a_1$ and $a_2$, were determined by comparison with solar-system experiments, 
while an upper 
bound was estimated for the $a_3.$\footnote{ For macroscopic matter, Nitsch and Hehl$^{20)}$ 
proposed a tetrad theory of gravitation as the translational guage limit of  
Poincar\'{e} 
gauge theory. Their choice of the parameters corresponds to $a_1=-1/3$, $a_2=1/3$, $a_3=3/2$ and 
$a_4=0$ in our notation.}$^)$ It was found that the numerical value of $a_1 +a_2$ should
 be very small, consistent with being zero. 

In the tetrad theory of gravitation, as far as we know, the total energy of an isolated system 
has been calculated only for spherically symmetric case. In the case $a_1+a_2=0=a_4$, Mikhail 
et al.$^{21)}$ found a static, spherically symmetric solution $({b^{k}}_{\mu})$ in Cartesian 
coordinates with $({b^{(0)}}_{\alpha}) \sim 1/\sqrt{r} \sim ({b^{a}}_{0})$ for 
r$\rightarrow \infty$, and showed that the total energy does not coincide with the gravitational 
mass. 
This result was extended to a wider class of solutions with spherical symmetry.$^{22)}$ An 
explicit expression was given for all the stationary, asymptotically flat solutions with 
${\it spherical}$ ${\it symmetry}$, which were then classified according to the asymptotic 
behavior of the components of 
$({b^a}_{0})$ and $({b^{(0)}}_{\alpha})$. It was found that the equality of the gravitational 
and inertial masses holds only when $({b^a}_{0})$ and $({b^{(0)}}_{\alpha})$ tend to zero faster
 than $1/{\sqrt{r}}$. 
 
 Calculation of the energy of an isolated system was extended to the generic case  
$(a_1+a_2)(a_1-4a_3/9)\neq 0$ under the assumption of {\it spherical} {\it symmetry}.$^{23)}$ 
It was shown that linear approximation can be applied to the field equation at large spatial 
distances from the source, and that the calculated energy is equal to the gravitational mass of
 the source. When $(a_1-4a_3/9)=0$, however, the problem of energy of an isolated system is not
 yet understood well,  even for the spherically symmetric case, since it is not always possible 
to  determine the asymptotic behavior of the components  $({b^a}_{\alpha})$ by means of  
linearized field equation.

It is the purpose of this paper to calculate the total energy and spatial momentum of an 
isolated system {\it
without} {\it assuming} {\it spherical} {\it symmetry}. For this purpose we solve the 
linearized 
 field equation at far distances up to order $O(1/r)$. The general solution has two constants. 
One of this is related to the gravitational mass of the source, and the other is a constant vector. We 
then show that the energy is always equal to the gravitational mass of the isolated system. In
 this sense the  equivalence principle is satisfied  in the tetrad theory of 
 gravitation. It should be noted that our conclusion is based on the assumption that weak 
field approximation can be applied at far distances from the source. In the spherically 
symmetric case mentioned above, however, solutions are known for which the linear approximation 
does not work even  far from the source because  the components, $({b^a}_{0})$ and 
$({b^{(0)}}_{\alpha})$, behave like $1/\sqrt{r}$.
 
 In \S 2 we discuss the most general Lagrangian with a parity-violating term and apply its 
field equation to the static, diagonal parallel vector fields with {\it spherical} 
{\it symmetry}. We find that the exact solution for the parity-conserving case$^{18)}$ 
satisfies the field equation.  In \S 3 we construct the linearized form of the field equation 
in vacuum at distances far from an isolated system. Assuming that the system is stationary,  
we obtain the general solution of the linearized field equation in vacuum. In \S 4 we derive 
the superpotential from the general Lagrangian and calculate its components necessary for
 computing the total energy and momentum of the system, using the solution obtained in \S 3. 
The final section is devoted to conclusion and  discussion.
 
\newsection{Basic Lagrangian}
%\newsection{Basic Lagrangian}
In a spacetime with absolute parallelism the parallel vector fields $({b^k}_{\mu})$ 
\footnote{Latin indices $(i,j,k,\cdots)$ designate the vector number, which runs from $(0)$ to
 $(3)$, while Greek indices $(\mu,\nu,\rho, \cdots)$ designate the world-vector components 
running from 0 to 3. The 
spatial part of Latin indices is denoted by $(a,b,c,\cdots)$, while that of Greek indices 
by $(\alpha, \beta,\gamma,\cdots)$.}$^)$ define the nonsymmetric connection 

\be
{\Gamma^\lambda}_{\mu \nu} ={b_k}^\lambda {b^k}_{\mu,\nu}
\ee
with ${b^k}_{\mu,\nu}= {\partial}_\nu {b^k}_{\mu}$, from which the torsion tensor is given by
\be
{T^\lambda}_{\mu \nu} = {\Gamma^\lambda}_{\mu \nu}-{\Gamma^\lambda}_{\nu \mu}
={b_k}^\lambda ({b^k}_{\mu,\nu}-{b^k}_{\nu,\mu}).
\ee
The curvature tensor defined by ${\Gamma^\lambda}_{\mu \nu}$ is identically 
vanishing, however. Here Latin indices are raised or lowered by the Minkowski 
metric 
$\eta_{i j}$=$\eta^{i j}$ =diag$(-1,+1,+1,+1)$. The metric tensor is given by the parallel vector fields 
as
\be
g_{\mu \nu}= b_{k \mu} {b^k}_ \nu.
\ee

Assuming  invariance under 
a) the group of general coordinate transformations, and 
b) the group of global Lorentz transformations, 
 we write the most general gravitational Lagrangian density quadratic in the torsion tensor as\footnote{
Throughout this paper we use the relativistic units, $c=G=1$. The Einstein constant ${\kappa}$ is then equal to $8 \pi $. We will denote the symmetric part by
( \ ), for example, $A_{(\mu \nu)}=(1/2)( A_{\mu \nu}+A_{\nu \mu})$
and the  antisymmetric part by the square bracket [\ ],
$A_{[\mu \nu]}=(1/2)( A_{\mu \nu}-A_{\nu \mu})$ .} 
\be
{\cal L}_G= {\sqrt{-g}\over \kappa} \left[a_1(t^{\mu \nu \lambda}
t_{\mu \nu \lambda})+a_2(v^{\mu} v_{\mu})+a_3(a^{\mu}a_{\mu}) +a_4(v^{\mu}a_{\mu})\right],
\ee
where $a_1$, $a_2$, $a_3$ and $a_4$ are 
dimensionless parameters of the theory, and
$t_{\mu \nu \lambda} $, $v_\mu$ and $a_\mu$ are the three irreducible components  of the torsion
 tensor.$^{18),}$\footnote{ The dimensionless parameters $\kappa a_i$ of Ref.$~$18) are here denoted by $a_i$ for convenience.} 

By applying the variational principle to the Lagrangian (4), we obtain the field 
equation:
\be
I^{\mu \nu}= {\kappa}T^{\mu \nu}
\ee
with
\be
I^{\mu \nu}=2{\kappa}[{D}_\lambda F^{\mu \nu \lambda}+ 
v_\lambda F^{\mu \nu \lambda}+H^{\mu \nu}
-{1 \over 2} g^{\mu \nu}L_G],
\ee
where
\be
F^{\mu \nu \lambda} =  {1 \over 2} b^{k \mu} {\partial L_G \over \partial 
{b^k}_{\nu,\lambda}} =-F^{\mu \lambda \nu},
\ee
\be
H^{\mu \nu} = T^{\rho \sigma \mu} 
 {F_{\rho \sigma}}^\nu - {1 \over 2} T^{\nu \rho \sigma} 
{F^\mu}_{\rho \sigma}=H^{\nu \mu},
\ee
\be
T^{\mu \nu} = {1 \over \sqrt{-g}} {\delta {\cal L}_M \over 
\delta {b^k}_\nu} b^{k \mu}.
\ee
Here ${L_G}= {{\cal L}_G/\sqrt{-g}}$, and ${\cal L}_M$ denotes the Lagrangian density of material fields, of which 
 the energy-momentum tensor $T^{\mu \nu}$ is nonsymmetric in general.

In ${\it static}$, ${\it spherically}$ ${\it symmetric}$ spacetime, the parallel vector fields take a diagonal form, and the field equation (5) 
can be exactly solved. 
The exact solution so obtained is the same as that obtained by assuming the invariance under 
parity operation.$^{18)}$ This implies that the parameter $a_4$ has no effect when we use diagonal parallel vector fields having spherical symmetry.

In order to reproduce the correct Newtonian limit, the parameters
$a_1$ and $a_2$ should satisfy the condition
\be
a_1+4a_2+9a_1a_2=0,
\ee
called the Newtonian approximation ${\rm condition,}^{18)}$  which can be 
solved to give
\be
a_1=-{1 \over 3(1-\epsilon)}, \quad 
a_2={1 \over 3(1-4\epsilon)},
\ee
in terms of a dimensionless parameter $\epsilon$. Comparison with 
solar-system experiments indicates that $|\epsilon|$ must be 
very small.

\newsection{Solution at far distances}
In order to calculate the energy and momentum of an isolated system confined in a finite 
spatial region, we focus our attention to the solution at far distances from the source without 
assuming 
spherical symmetry. In such spatial regions far from the source, the gravitational field is
 weak and matter fields are not present: Thus, we are allowed to treat the linearized field 
equation in vacuum.  For simplicity we assume that the whole spacetime is covered with a single
coordinate system 
\{$x^{\mu}$; $+\infty>x^{\mu}>-\infty$\} with the origin being located  somewhere inside the 
finite system. In spatial regions far from the source, the parallel vector fields can be 
represented as
\ba
{b^k}_{\mu}(x) \A = \A {\delta^k}_{\mu}+{a^k}_{\mu}, \quad |{a^k}_{\mu}|<<1,\\
{b_k}^{\mu}(x) \A = \A {\delta_k}^{\mu}+{c_k}^{\mu}, \quad |{c_k}^{\mu}|<<1,
\ea
and we  assume that all the quadratic and higher-order terms of ${a_k}^{\mu}$ 
(or ${c_k}^{\mu}$) can be neglected in the field equation. 
Accordingly, we do not distinguish Latin indices from Greek indices 
for ${a^k}_{\mu}$ and ${c_k}^{\mu}$: We use Greek indices which are now raised and lowered by
 the Minkowski metric  $\eta^{\mu \nu}$ and $\eta_{\mu \nu}$. 
>From the  relation  ${b^k}_{\mu} {b_k}^{\nu}={\delta_\mu}^{\nu}$, it 
 follows that
\be
a_{\nu \mu}+c_{\mu \nu}=0
\ee
with $a_{\nu \mu}=\eta_{\nu \lambda} {a^{\lambda}}_{\mu}$ and $c_{\mu \nu}=\eta_{\nu \lambda} {c_\mu}^{\lambda}$.
We decompose $a_{\mu \nu}$ into symmetric and antisymmetric parts,
\be
a_{\mu \nu}={1 \over 2}h_{\mu \nu}+A_{\mu \nu},
\ee
with $h_{\mu \nu}=h_{\nu \mu}$ and  $A_{\mu \nu}=-A_{\nu \mu}$. The components of the metric tensor are written as 
\be
g_{\mu \nu}=\eta_{\mu \nu}+h_{\mu \nu}.
\ee
The antisymmetric part has no contribution to the spacetime metric, implying that it is associated with the intrinsic spin-${1/2}$ fundamental particles. 

Keeping only the linear terms and putting $T_{\mu \nu}=0$ on the right-hand side, we see that the symmetric part and the skew-symmetric part of the field equation (5) take the form
\ba
{3a_1 \over 2}\Box{\overline h}_{\mu \nu}-
{1 \over 2}(a_1+a_2)(\eta_{\mu \nu} \Box{ \overline h} - 
{\overline h}_{,\mu \nu})+({a_1 \over 2} -a_2)
\eta_{\mu \nu}{{\overline h}^{\rho \sigma}}_{,\rho \sigma} \nonu
-(2a_1-a_2) {{\overline h}^{\rho}}_{(\mu, \nu) \rho} 
+2(a_1+a_2){A^\rho}_{(\mu, \nu) \rho} 
+{2 \over 3}a_4{{\overline A}^{\rho}}_{(\mu, \nu) \rho}=0, 
\ea
and 
\ba
-(a_1+a_2){\overline h^\rho}_{[\mu, \nu] \rho}+
(a_1-{4 \over 9}a_3) \Box A_{\mu \nu}-
2(a_2+{4 \over 9}a_3){A^{\rho}}_{[\mu, \nu] \rho} \nonu
-{2 \over 3}a_4 {{\overline A}^{\rho}}_{[\mu, \nu] \rho} 
+{1 \over 3}a_4 
\epsilon_{\mu \nu \lambda \rho} \partial^{\lambda} \partial_
{\sigma}({1 \over 2} {{\overline h}^{\rho \sigma}}-A^{\rho \sigma})
=0,
\ea
respectively, in spatial regions far from the source. Here the d'Alembertian operator is 
given by
$\Box = \partial^{\mu} \partial_{\mu}$,  ${\overline h}_{\mu \nu}$ denotes 
\be
{\overline h}_{\mu \nu}=h_{\mu \nu}-{1 \over 2}\eta_{\mu \nu}h, \qquad 
h=\eta_{\mu \nu}h^{\mu \nu},
\ee
and ${\overline A}_{\mu \nu}$ stands for the dual of $A_{\mu \nu}$ defined by
\be
{\overline A}_{\mu \nu}={1 \over 2}\epsilon_{\mu \nu \lambda \rho}
A^{\lambda \rho},
\ee
where ${\epsilon}_{\mu \nu \rho \sigma}$ is the completely antisymmetric 
tensor normalized as ${\epsilon}^{0123}=+1$.
It is clear from (17) and (18) that the 
symmetric 
field $h_{\mu \nu}$ and the skew-symmetric field $A_{\mu \nu}$ are coupled to each other 
unless the parameters satisfy the condition $a_1+a_2=0$ and $a_4=0$. 
As is easily checked, the linearized field equations
 (17) and (18) are invariant under the gauge transformation
\be
{{h^{\prime}}}_{\mu \nu}=h_{\mu \nu}-2\xi_{(\mu,\nu)},\quad 
{{A^{\prime}}}_{\mu \nu}=A_{\mu \nu}-\xi_{[\mu,\nu]},
\ee
where the $\xi_{\mu}$  are  small functions which leave the fields weak. By means of this 
freedom we can require the gauge condition 
\be
\partial_{\nu}{\overline h}^{\mu \nu}=0,
\ee
which we shall assume henceforth. Then Eqs. (17) and (18) become
\be
{3a_1 \over 2} \Box {\overline h}_{\mu \nu}-
{1 \over 2}(a_1+a_2)(\eta_{\mu \nu}\Box{\overline h}- {\overline h}_{,\mu \nu})+2(a_1+a_2){A^\rho}_{(\mu, \nu) \rho} 
+{2 \over 3}a_4{{\overline A}^{\rho}}_{(\mu, \nu) \rho} =0,
\ee
\be
(a_1-{4 \over 9}a_3) \Box A_{\mu \nu}-
2(a_2+{4 \over 9}a_3){A^{\rho}}_{[\mu, \nu] \rho}
-{2 \over 3}a_4 {{\overline A}^{\rho}}_{[\mu, \nu] \rho} 
-{1 \over 3}a_4 
\epsilon_{\mu \nu \lambda \rho} {\partial}^{\lambda}{\partial}_{\sigma}
{A^{\rho \sigma}}=0.
\ee

In order to calculate the total energy and momentum of the isolated system under consideration, it is enough to obtain asymptotic solutions of (23) and (24) up to order $O(1/r)$. Multiplying $\partial_{\nu}$ on (23) [or (24)], we obtain
\be
\Box \left[ (a_1+a_2){A^\nu}_{\mu}+
{\tilde a_4} {{\overline A}^{\nu}}_{\mu} \right]_{,\nu}=0 
\ee
with ${\tilde a_4}=(a_4/3)$. Let us define $B_{\mu \nu}
=B_{[\mu \nu]}$ by
\be
B_{\mu \nu}=(a_1+a_2)A_{\mu \nu}+{\tilde a_4}{\overline A}_{\mu \nu},
\ee
which can easily be solved with respect to $A_{\mu \nu}$ to give
\be
A_{\mu \nu}={1 \over (a_1+a_2)^2+{\tilde a_4}^2}
\left[ (a_1+a_2)B_{\mu \nu}-{\tilde a_4}{\overline B}_{\mu \nu} \right],
\ee
under the assumption that $(a_1+a_2)^2+{\tilde a_4}^2\neq0$, or equivalently 
$(a_1+a_2)\neq0$ and/or $a_4\neq0$. We note that when this assumption is violated, the 
theory is reduced to the special case of Hayashi and Nakano$^{16)}$ and M\o ller.$^{17)}$

We now assume that the system under consideration is stationary and that the system as a whole is at rest somewhere around the origin. Then $h_{\mu \nu}$ and $A_{\mu \nu}$ are time-independent. 
Rewriting (23) and (24) in terms of $B_{\mu \nu}$, we have after a slight modification
\be
\Delta {\overline h} =0,
\ee
\be
{3a_1 \over 2} \Delta {\overline h}_{0 0}=0,
\ee
\be
{3a_1 \over 2} \Delta {\overline h}_{0 \alpha}-B_{0 \beta ,\alpha \beta}=0,
\ee
\be
f_1 \Delta B_{0 \alpha}+f_2 B_{0 \beta, \beta \alpha}
+f_3 \epsilon_{\alpha \beta \gamma} B_{\delta [\beta,\gamma] \delta}=0, 
\ee
\be
{3a_1 \over 2} \Delta {\overline h}_{\alpha \beta}+{1 \over 2}
(a_1+a_2){\overline h}_{,\alpha \beta}+
2B_{\gamma (\alpha,\beta) \gamma}=0,
\ee
\be
f_1 \Delta B_{\alpha \beta}-2f_2 B_{\delta [\alpha, \beta] \delta}
+f_3 \epsilon_{\alpha \beta \gamma} B_{0 \delta,\delta \gamma}=0, 
\ee
where we have introduced $f_1$, $f_2$ and $f_3$ by
\ba
f_1 \A = \A (a_1+a_2)(a_1-\displaystyle {4 \over 9}a_3)+{\tilde a_4}^2,\\
f_2 \A = \A  (a_1+a_2)(a_2+\displaystyle{4 \over 9}a_3),\\
f_3 \A= \A {\tilde a_4}(a_2+\displaystyle{4 \over 9}a_3).
\ea
We note that except for $\Delta B_{0 \alpha}$ and $\Delta B_{\alpha \beta}$ Eqs. (30)$\sim$(33) involve $B_{0 \alpha}$ and 
$B_{\alpha \beta}$ only in the form of 3-dimensional divergence.

According to (25) with (26) the divergences, $B_{0 \alpha, \alpha}$ and 
$B_{\alpha \beta, \beta}$, satisfy the Laplace equation, and therefore, they are  given up 
to $O(1/r^2)$ by
\ba
B_{0 \alpha,\alpha} \A=\A {\grave B_\alpha n_\alpha \over r^2},\\
B_{\alpha \beta,\beta} \A=\A {\grave F_{\alpha \beta} n_\beta \over r^2},
\ea
where ${\grave B_\alpha}$ is a constant vector, and ${\grave F_{\alpha \beta}}$ is a nonsymmetric constant tensor. Here the radial unit vector $n^\alpha$ is defined by 
$n^\alpha=\displaystyle{x^\alpha/r}$ 
without making distinction between upper and lower indices. We have omitted from our 
consideration the possibility that the leading terms of the solutions (37) and (38) begin 
from $O(1/r)$. 
This is because if, on the contrary, they begin from $O(1/r)$-terms, Eqs. (30)$\sim$(33) 
will have no solutions of ${\overline h}_{\mu \nu}$ and $B_{\mu \nu}$  which tend to zero for 
large $r$. 
Since $B_{\alpha \beta}$ is by definition antisymmetric with respect to $\alpha$ and $\beta$, the constant tensor ${\grave F_{\alpha \beta}}$ must be of the form
\be
{\grave F_{\alpha \beta}}={\grave F_{[\alpha \beta]}}+{1 \over 3}
{\grave F}\delta_{\alpha \beta}
\ee
with ${\grave F}$ being the trace. 
With the help of (37)$\sim$(39) we can solve the field equations (28)$\sim$(33) by 
 following the ordinary procedure to solve the Laplace equation. (See the Appendix for a summary of the procedure.)

>From (28) we see that ${\overline h}$ can be expressed by  ${\overline h}=p/r$ up to $O(1/r)$, where $p$ is a constant. Using this expression for ${\overline h}$ and Eq. (38) in (32), we obtain
\be
{\overline h_{\alpha \beta}}={\grave G_{\alpha \beta} \over r}+
{1 \over 6a_1} \left\{ 4 \delta_{\delta (\alpha}{\grave F_{\beta) \gamma}}
-{4 \over 9}{\grave F}\delta_{\alpha \beta} \delta_{\gamma \delta}
+ p(a_1+a_2) \left( \delta_{\alpha \gamma} \delta_{\beta \delta} 
 -{1 \over 3}\delta_{\alpha \beta} \delta_{\gamma \delta} \right) \right\} 
 {n_\gamma n_\delta \over r},
 \ee
 where ${\grave G_{\alpha \beta}}$ is a symmetric, constant tensor. 
Applying the condition (22) to ${\overline h}_{\alpha \beta}$ of (40), we see that  ${\grave G_{\alpha \beta}}$ is given by 
\be
{\grave G_{\alpha \beta}}={2 \over 3a_1} \left[ {4 \over 9}{\grave F}
\delta_{\alpha \beta}+{\grave F_{[\alpha \beta]}}+{1 \over 3}(a_1+a_2) p \delta_{\alpha \beta} \right].
\ee
Since ${\grave G_{\alpha \beta}}$ is symmetric, Eq. (41) implies
\be
{\grave F_{[\alpha \beta]}}=0.
\ee

As for the field equation for $B_{\alpha \beta}$ we note that the second term on the left-hand side of (33) vanishes owing to (39) and (42). 
Applying the same procedure to (33), we obtain
\be
B_{\alpha \beta}={\grave H_{\alpha \beta} \over r}+
{f_3 \over 6f_1} {\epsilon_{\alpha \beta \gamma} \over r}
({\grave B_\gamma} 
-3 n_\gamma n_\delta {\grave B_\delta}) ,
\ee
 where we have assumed that the constant $f_1$ of (34) is nonvanishing. 
Here ${\grave H}_{\alpha \beta}$ is an antisymmetric constant tensor to be fixed by the relation (38): We have 
\be
{\grave H_{\alpha \beta}}={f_3 \over 3f_1} \epsilon_{\alpha \beta \gamma}
{\grave B_\gamma},
\ee
\be
{\grave F}=0,
\ee
the latter of which together with (39) and (42) implies that 
$B_{\alpha \beta,\beta}$ vanishes up to $O(1/r^2)$.

We can analyze the field equations for ${\overline h_{0 \alpha}}$ and 
$B_{0 \alpha}$ in the same manner. We here summarize the asymptotic solution up to 
$O(1/r)$ for ${\overline h_{\mu \nu}}$  and $B_{\mu \nu}$ as follows:
\be
{\overline h}={p \over r},
\ee
\be
{\overline h}_{0 0}={(2a_2-a_1)p \over 3a_1 r},
\ee
\be
{\overline h}_{0 \alpha}={{\grave B_\alpha}+n_\alpha n_\beta{\grave B_\beta}  \over 3a_1 r},
\ee
\be
{\overline h}_{\alpha \beta}={(a_1+a_2)p \over 6a_1 r}(\delta_{\alpha \beta}+n_\alpha n_\beta),
\ee
\be
B_{0 \alpha}=-{1 \over r} \left[{\grave B_\alpha}+{f_2 \over 2f_1}
\left( {\grave B_\alpha}+n_\alpha n_\beta{\grave B_\beta} \right) 
 \right],
\ee 
\be
B_{\alpha \beta}={f_3 \over 2f_1} 
{\epsilon_{\alpha \beta \gamma} \over r}({\grave B_\gamma} 
-n_\gamma n_\delta{\grave B_\delta} ),
\ee 
where $p$ is an unknown constant and ${{\grave B}_\alpha}$ is an unknown constant vector. From (19) and (27), we obtain the most general expression for $h_{\mu \nu}$ and $A_{\mu \nu}$ up to $O(1/r)$
\be
h =-{p \over r},
\ee
\be
h_{0 0}=-{3a_2 p \over 2 r},
\ee
\be
h_{0 \alpha} ={{\grave B_\alpha}+n_\alpha n_\beta{\grave B_\beta}  
\over 3a_1 r}, 
\ee
\be
h_{\alpha \beta} =-{p \over 2 r}\delta_{\alpha \beta}+{p (a_1+a_2) \over 6a_1}
{(\delta_{\alpha \beta}+ n_\alpha n_\beta) \over r},
\ee
\be
A_{0 \alpha}=-{1 \over 2f_1 r} \left[2{\grave B_\alpha} \left(a_1-\displaystyle{4 
\over 9}a_3 \right)+\left(a_2+\displaystyle{4 \over 9}a_3 \right) 
\left( {\grave B_\alpha} +n_\alpha n_\beta {\grave B_\beta} \right)  \right], 
\ee 
\be
A_{\alpha \beta}={{\tilde a_4} \over f_1 r } {\epsilon}_{\alpha \beta \gamma}{\grave B}_\gamma.
\ee 
The covariant components of the parallel vector fields are then represented up to $O(1/r)$ by
\be
{b^{(0)}}_{0} = 1-{m \over r},
\ee
\be
{b^{(0)}}_{\alpha} ={\left(a_1-\displaystyle{4 \over 9}a_3\right) \over f_1 }
{{\grave B_\alpha} \over r}
 -\left \{ {1 \over 3a_1}-{ \left(a_2+\displaystyle{4 \over 9}a_3\right) \over f_1} \right
  \} {\left( {\grave B_\alpha} + n_\alpha n_\beta {\grave B_\beta} \right) 
\over 2r},
\ee 
\be
{b^{(a)}}_{0} ={\left(a_1-\displaystyle{4 \over 9}a_3\right) \over 
f_1 }{{\grave B_a} \over r}
 +\left \{ {1 \over 3a_1}+{ \left(a_2+\displaystyle{4 \over 9}a_3\right) \over f_1} 
 \right \}
{ \left( {\grave B_a} + n_a n_\beta{\grave B_\beta} \right) \over 2r},
\ee 
\be
{b^{(a)}}_{\alpha}=\delta_{a \alpha}-{m \left \{ (a_2-2a_1) \delta_{a \alpha}+
(a_1+a_2) n_a n_\alpha \right \} \over 9a_1 a_2 r} 
+{{\tilde a_4}{\epsilon}_{a \alpha \beta}{\grave B}_\beta \over f_1 r } 
\ee
with {\it m} being a constant given by
\be
m=-{3a_2 p \over 4}.
\ee
Thus the parallel vector fields are asymptotically characterized by the constant vector 
${\grave B}_{\alpha}$ besides the constant $m$ at large spatial distances from an isolated, 
stationary system. The constant $m$ can be interpreted as the gravitational mass of the source,  
as  seen from (53). The physical meaning of the constant ${\grave B}_{\alpha}$ will become 
clear when we calculate in the next section the total energy and momentum of the isolated 
system.

The parallel vector fields of (58)$\sim$(61) are not spherically symmetric because they involve
 the constant vector ${\grave B}_{\alpha}$. This is to be contrasted  with the metric in general 
relativity, which is governed by the Einstein equation.$^{24)}$ More precisely, if one 
selects a system of coordinates with respect to which the isolated system is as a whole at rest, 
and if one neglects gravitational radiation by the system, the Einstein equation requires that 
the metric is spherically symmetric and static up to $O(1/r)$ for large $r$. We note, however, 
that when ${\grave B}_{\alpha}=0$, the parallel vector fields (58)$\sim$(61) reduce to the 
static, spherically symmetric solution expanded up to $O(1/r)$.

We have been assuming that the parameters  satisfy $(a_1+a_2)^2+{\tilde a_4}^2\neq0$ and 
$f_1\neq0$. Now that we have obtained the solution of 
$({b^k}_{\mu})$ up to $O(1/r)$, we can study the limiting cases  $a_4=0\neq(a_1+a_2)$ and 
$(a_1+a_2)=0\neq a_4$. In both cases we can use $A_{\mu \nu}$ directly instead of $B_{\mu \nu}$.
 In the case  $a_4=0$ and $(a_1+a_2)\neq0$, we can further take the limit $(a_1+a_2)\rightarrow0$: It is easy to check that the resultant $h_{\mu \nu}$ and $A_{\mu \nu}$ satisfy Eqs. (23) and (24) with $(a_1+a_2)=0$ and $a_4=0$. 
 In  case  $(a_1+a_2)=0$ and $a_4\neq 0$, on the other hand, it is possible  to take 
  limit  $a_4 \rightarrow 0$ only when $(a_1-4a_3/9) \rightarrow 0$ is taken 
simultaneously. The resultant theory is just the tetrad formulation of general relativity. 

\newsection{Calculation of energy and momentum}

Since the total Lagrangian is a scalar under general coordinate transformations, the total energy-momentum complex $({T_\mu}^\nu$+${t_\mu}^\nu)$ is represented  as 
\be
\sqrt{-g}({T_\mu}^\nu+{t_\mu}^\nu)={{{\cal U}_\mu}^{\nu \lambda}}_{,\lambda}
\ee
with the superpotential ${{{\cal U}_\mu}^{\nu \lambda}}$ being antisymmetric in $\nu$ and $\lambda$. Here ${t_\mu}^{\nu}$ is the canonical energy-momentum complex for the gravitational field derived from the gravitational Lagrangian density (4).$^{22)}$ The superpotential ${{{\cal U}_\mu}^{\nu \lambda}}$ is given by 
\ba
{{{\cal U}_\mu}^{\nu \lambda}} \A =\A 2\sqrt{-g}{F_\mu}^{\nu \lambda}\nonu
%
%\A \A \quad
\A = \A {2 \sqrt{-g} \over \kappa} \Biggl[ 
\left( a_1-{a_3 \over 9} \right) {T_\mu}^{\nu \lambda}
+\left( {a_1 \over 2}+{a_3 \over 9} \right)
 \left( {T^{\lambda \nu}}_{\mu}-{T^{\nu \lambda}}_\mu \right) 
-\left( {a_1 \over 2} - a_2 \right)
\left( {\delta_\mu}^\nu v^\lambda-{\delta_\mu}^\lambda v^\nu \right) \nonu
\A \A \quad
+{a_4 \over 12} \left(
{\delta_\mu}^{\nu} \epsilon^{\lambda \rho \sigma \tau}T_{\rho \sigma \tau}-{\delta_\mu}^{\lambda} \epsilon^{\nu \rho \sigma \tau}T_{\rho \sigma \tau} -2{\epsilon_\mu}^{\nu \lambda \rho}v_\rho \right) \Biggr],
\ea
which satisfies the M\o ller condition (C).  Since we have solved the field equation 
approximately at far distances, it is better to rewrite the superpotential (64) in the asymptotic form 
\newpage
\ba
{{\cal U}_\mu}^{\nu \lambda} \A = \A {2 \over \kappa} \Biggl [
2 \left( a_1-{a_3 \over 9} \right){{b_{\mu}}^{[\nu,\lambda]}}
%+\left(a_1+{2 a_3 \over 9} \right)%
% \left( {b^{[\lambda \nu]}}_{,\mu}-{\partial}^{[\nu}%
% {b^{\lambda]}}_{\mu} \right) %
 -\left(a_1  - 2a_2 \right)
\left( {\delta_{\mu}}^{[\nu}{\partial}^{\lambda]} {b^{\rho}}_{\rho}-
{b^{\sigma [\lambda}}_{,\sigma} {\delta^{\nu]}}_{\mu} \right) \nonu
 %\left(a_1  - 2a_2 \right)%
%\left( {\delta_{\mu}}^{[\nu}{\partial}^{\lambda]} {b^{\rho}}_{\rho}-%
%{b^{\sigma [\lambda}}_{,\sigma} {\delta^{\nu]}}_{\mu} \right) \nonu%
%
\A \A 
+\left(a_1+{2 a_3 \over 9} \right)
 \left( {b^{[\lambda \nu]}}_{,\mu}-{\partial}^{[\nu}
 {b^{\lambda]}}_{\mu} \right) 
%-\left(a_1  - 2a_2 \right)%
%\left( {\delta_{\mu}}^{[\nu}{\partial}^{\lambda]} {b^{\rho}}_{\rho}-%
%{b^{\sigma [\lambda}}_{,\sigma} {\delta^{\nu]}}_{\mu} \right)%
+{a_4 \over 3} \left \{{\delta_\mu}^{[\nu} \epsilon^{\lambda] \rho \sigma \tau}
{b_{\rho \sigma ,\tau}} +{\epsilon_{\mu}}^{\nu \lambda \rho}
{b^{\sigma}}_{[\rho,\sigma]}
 \right\} \Biggr ], 
\ea
keeping only $O(1/r^2)$-terms. Here we use only Greek indices and raise (or lower) them with 
the Minkowski metric: Thus, for example, ${b^{\lambda}}_{\mu}=
{{\delta^{\lambda}}_k}{b^k}_{\mu}$ and 
${b_{\mu}}^{\nu,\lambda}={\eta^{\lambda \rho}}{\partial}_{\rho}{b_{\mu}}^{\nu}$. 
he formula for the energy and spatial momentum are given ${\rm by}^{7)}$
\be
E=- \lim_{r \rightarrow \infty}\int_{\rm r=const} {{\cal U}_{0}}^{0 \alpha} dS_{\alpha},
\ee
\be
P_{\alpha}=\lim_{r \rightarrow \infty}\int_{\rm r=const} 
{{\cal U}_{\alpha}}^{0 \beta} dS_{\beta}.
\ee

The gravitational Lagrangian is assumed to be a quadratic invariant of the torsion tensor, and accordingly the canonical complex ${t_\mu}^\nu$ behaves as $O(1/r^4)$ at far distances since the parallel vector fields are asymptotically given by (58)$\sim$(61). Thus, the M\o ller condition (B) is also satisfied, and the energy and momentum are transformed as a 4-vector. 

Now let us calculate the energy by using the asymptotic form of the parallel vector fields. The necessary components of the superpotential  are given by
\ba
{{\cal U}_0}^{0 \alpha}\A =\A {2 \over \kappa r^2} \left [ \left \{ 
(a_1+a_2)+{(2a_2-a_1)^2 \over 9a_1 a_2} \right \} m n_{\alpha} 
+{3a_1{\tilde a_4} \over 2f_1} \epsilon_{\alpha \beta \gamma} n_{\beta}{\grave B}_{\gamma} \right] \nonu
\A =\A -{2 \over \kappa r^2} \left ( m n_{\alpha} 
-{3a_1{\tilde a_4} \over 2f_1} \epsilon_{\alpha \beta \gamma} n_{\beta}{\grave B}_{\gamma} \right),
\ea
where we have used the Newtonian approximation condition (10) to rewrite the first term inside the square bracket. It is easily shown that Eq. (68) satisfies 
${{{\cal U}_0}^{0 \alpha}}_{,\alpha}=0$ up to $O(1/r^3)$ as it should. 
Using (68) in 
(66) we find that the last term of (68) does not contribute to the integral and  that the energy is given by
\be
E=m.
\ee
This  shows that the most general Lagrangian (4) including a parity-violating term  is consistent with the equivalence principle. 

Next, we turn to the spatial momentum $P_{\alpha}$. The necessary components of the superpotential are given by
\newpage
\ba
{{\cal U}_{\alpha}}^{0 \beta} \A = \A {2 \over \kappa f_1 r^2} \Biggl [ 
\left(a_1-\displaystyle{4a_3 \over 9} \right) \left \{{3 a_1 \over 2}n_{\beta}
{\grave B_{\alpha}}
-{1 \over 2} (a_1-2a_2)\delta_{\alpha \beta}n_{\gamma}{\grave B_{\gamma}}\right \} \nonu
\A \A 
+2\left(a_1+{2a_3 \over 9} \right)(a_1+a_2)n_{[\alpha} {\grave B_{\beta]}} -{{\tilde a_4} \over 6a_2}f_1 m n_{\rho}{\epsilon}_{\alpha \beta \rho}- {\tilde a_4}^2 \left( n_{[\alpha}{\grave B_{\beta]}}-\delta_{\alpha \beta} n_{\gamma}{\grave B_{\gamma}} \right) \nonu
\A \A 
%- {\tilde a_4}^2 \left( n_{[\alpha}{\grave B_{\beta]}}-\delta_{\alpha \beta}%
% n_{\gamma}{\grave B_{\gamma}} \right)%
-{1 \over 4} \left(f_1+3a_1\left(a_2+{4a_3 \over 9} \right) \right ) \left \{ 2n_{[\alpha}{\grave B_{\beta]}}
+(\delta_{\alpha \beta}-3n_{\alpha}n_{\beta})n_{\gamma}{\grave B_{\gamma}}
\right \} \Biggr ].
\ea
Taking the divergence of (70), we find that 
${{{\cal U}_{\alpha}}^{0 \beta}}_{,\beta}=0$, which means that the leading term of the 
${{{\cal U}_{\alpha}}^{0 \beta}}_{,\beta}$ is of $O(1/r^4)$, as it should.   Now we are ready to 
calculate the spatial momentum to determine  if it is vanishing  or not. Using (70) in (67), we find 
after a lengthy calculation
\be
P_{\alpha}={\grave B_{\alpha}},
\ee
which shows that the constant vector ${\grave B}_{\alpha}$ has the physical meaning of the 
spatial momentum of the isolated system. 

Since the isolated system is assumed to be stationary and as a whole at rest near  the origin, 
its spatial momentum should be vanishing, and hence the constant vector ${\grave B}_{\alpha}$ 
must be zero. This means that the divergence $B_{0 \alpha,\alpha}$ should be at most of 
$O(1/r^3)$ for  parallel vector fields at distances far from a stationary, isolated system. 
As we have seen in the previous section, however, such an  asymptotic behavior of  
$B_{0 \alpha,\alpha}$ does not follow from the linearized field equation in vacuum. Accordingly 
the asymptotic condition,
\be
\lim_{r \rightarrow \infty} B_{0 \alpha,\alpha}=O(1/r^3),
\ee
must be imposed by hand, in addition to the condition (12) (or (13)). By contrast, as we noted below (45), the linearized field equation implies the condition 
$B_{\alpha \beta,\beta}=O(1/r^3)$.

The asymptotic condition (72) is indeed satisfied when the gravitational field is weak 
everywhere and the linearized field equation, which is just given by Eqs.(17) and (18) with 
the right-hand sides being replaced by $\kappa T_{(\mu \nu)}$ and $\kappa T_{[\mu \nu]}$, 
respectively, is valid throughout whole space. Here $T_{\mu \nu}$ is the energy-momentum tensor
 in the special relativistic limit and satisfies the conservation law ${\partial}_{\nu} T^{\mu \nu}=0$. Multiplying 
 both sides of the linearized field equation by ${\partial}^{\nu}$, and using the conservation law, we obtain the inhomogeneous Laplace equation for $B_{0 \alpha,\alpha}$
\ba
\Delta B_{0 \alpha,\alpha} \A =\A \kappa T_{[0 \alpha],\alpha} \nonu
%
%\A \A \quad
\A =\A {\kappa \over 2} S_{0 \alpha \beta, \alpha \beta},
\ea
where $S^{\mu \nu \lambda}$ is the intrinsic spin tensor of the source, and we have used the 
Tetrode formula $T^{\mu \nu}=(1/2){\partial}_{\lambda} S^{\mu \nu \lambda}$ in the last step.
 For an isolated system confined to a finite spatial region, the solution $B_{0 \alpha,\alpha}$ 
of (73) is of $O(1/r^3)$ for large $r$.

In the parity-conserving case with $a_4=0$ it is more convenient to use $A_{0 \alpha}$ and $A_{\alpha \beta}$ 
directly. Writing the asymptotic form of $A_{0 \alpha,\alpha}$, which satisfies the Laplace equation, as $B_\alpha n_\alpha/r^2$, Eq. (71) reads 
\be
P_{\alpha}=(a_1+a_2){B_{\alpha}}.
\ee
As explained at the end of \S 3, the solution of $h_{\mu \nu}$ and $A_{\mu \nu}$ has a well-defined limit when we put $a_4=0$ and then $(a_1+a_2)=0$. Although the resultant solution involves the constant vector $B_\alpha$, the total spatial momentum is vanishing as is shown in (74). This situation can be understood as follows. 
 In the special case  $(a_1+a_2)=0=a_4$, the linearized field equation for $A_{\mu \nu}$ 
is decoupled from that for $h_{\mu \nu}$, and therefore, the former equation is invariant under 
the gauge transformation
\be
{A^{\prime}}_{\mu \nu}=A_{\mu \nu}+\zeta_{[\mu,\nu]}
\ee
with $\zeta_{\mu}$ representing arbitrary small functions. We can use this  freedom to choose 
a gauge in which $A_{0 \alpha}$ and $A_{\alpha \beta}$ are vanishing up to $O(1/r)$. Thus, 
in this special case, the constant vector ${\grave B}_{\alpha}$ is unphysical  and can be eliminated by a gauge transformation.

\newsection{Conclusion and discussion}
%{\large {\bf Conclusion and discussion}}

We have studied the energy and momentum of an isolated system in the tetrad theory of 
gravitation, starting from the most general Lagrangian that is quadratic in torsion and involves four 
unknown parameters $a_1$, $a_2$, $a_3$ and $a_4$, the last of which is associated with a 
parity-violating term. As a first application we considered the static, spherically symmetric 
case, where the parallel vector fields take a diagonal form. The solution of the field equation 
in vacuum is found to be the same as the exact solution of the parity-conserving case with 
$a_4=0$.$^{18)}$ This is due to the fact that the axial-vector part  of the torsion tensor is
 identically vanishing  for  diagonal parallel vector fields.

The total energy and momentum of an isolated system are expressed by a surface integral over a 
large closed surface enclosing the system. It is then sufficient to know the asymptotic form of 
the parallel vector fields at distances far from the source, where the gravitational field is 
weak. In view of this, we analyzed the linearized field equation in vacuum which follows from the most
 general gravitational Lagrangian. 

It is well known that in general relativity the Einstein equation ensures that the metric tensor
 at far distances is the same as the Schwarzschild metric up to  $O(1/r)$ for any isolated 
stationary system. This is usually shown$^{24)}$ by solving the linearized Einstein equation in 
vacuum, which  is the Laplace equation in stationary case and can easily be solved.   
 
 In our case, the linearized field equation in vacuum consists of 16 equations for the symmetric 
field $h_{\mu \nu}$ and the skew-symmetric field $A_{\mu \nu}$,   and is invariant under a gauge 
transformation, which allows us to impose the harmonic condition on the $h_{\mu \nu}$. It is found,
 however, that the symmetric part and skew-symmetric part of the linearized field equation are 
coupled with each other unless the parameters satisfy $a_1+a_2=0=a_4$. Nevertheless, we can 
solve the coupled equation  up to  $O(1/r)$ assuming that the parameters satisfy 
$(a_1+a_2)^2+(a_4/3)^2\neq 0$ and $(a_1+a_2)(a_1-4a_3/9)+(a_4/3)^2\neq0$. It is found that 
the general solution has two constants, the gravitational mass $m$ and a constant vector 
${\grave B}_{\alpha}$. The general solution  of the linearized field equation up to $O(1/r)$ is,
 therefore, not spherically symmetric, due to the precense of ${\grave B_\alpha}$. 

  Using the definition of energy-momentum complex given by M\o ller,$^{12)}$ we derive the 
superpotential ${{\cal U}_{\mu}}^{\nu \lambda}$ from the general Lagrangian. It transforms as a 
tensor density under general coordinate transformations. Using the general solution obtained, 
we give an explicit expression for  the components ${{\cal U}_{0}}^{0 \alpha}$ and 
${\cal U_\alpha}^{0 \beta}$, 
    which are shown to be divergenceless up to $O(1/r^3)$, implying that the leading term of the 
divergence of those components is of order $O(1/r^4)$. 

  From the components ${{\cal U}_{0}}^{0 \alpha}$ we calculate the total energy, and show that
 the result is equal to the gravitational mass of the isolated system, i.e. $E=m$.
 Using the components ${\cal U_\alpha}^{0 \beta}$, we then calculate the spatial momentum, and 
find that  $P_\alpha={\grave B_\alpha}$. Thus, we arrive at the result that the general 
solution at far distances is characterized by the total energy and the spatial momentum of the
isolated system under consideration.

The asymptotic solution with vanishing ${\grave B_\alpha}$ is acceptable as a description of the 
parallel vector fields far from a stationary isolated system which is at rest as a whole. It
indicates that the total momentum of the isolated system at rest is vanishing. The asymptotic 
 solution coincides with the exact solution with spherical symmetry up to $O(1/r)$.

It is not yet clear whether or not the asymptotic solution with ${\grave B_\alpha} \neq 0$,   
actually 
describes the asymptotic behavior of exact solutions. Such exact solutions, if they exist, should 
 describe exotic systems which have nonvanishing total momentum although at rest as a whole.

The vanishing of the constant vector ${\grave B_\alpha}$ implies that the divergence 
$B_{0 \alpha,\alpha}$
 of $B_{0 \alpha}=(a_1+a_2)A_{0 \alpha}+(a_4/3){\overline A}_{0 \alpha}$ 
behaves at most as $O(1/r^3)$ for large $r$. This asymptotic behavior does not follow from 
 the linearized field equation alone. By contrast, the asymptotic behavior of 
$B_{\alpha \beta}=(a_1+a_2)A_{\alpha \beta}+(a_4/3){\overline A}_{\alpha \beta}$ is 
more severely governed by 
the linearized field equation: In fact, it is shown that  $B_{\alpha \beta,\beta}$
 vanishes up to $O(1/r^2)$.

For a weakly gravitating source, for which the field is weak everywhere and the weak field 
approximation can be applied, we can show that  ${B_{o \alpha,\alpha}}$ behaves as  
$O(1/r^3)$. For the fully relativistic case we do not know the reason why  
${B_{o \alpha,\alpha}}$  should behave in this manner. Therefore, further study 
is required to establish the asymptotic condition of the parallel vector 
fields, in particular, of the skew-symmetric field $A_{\mu \nu}$. 
 
The study of the weak field case also needs more investigation.  Although in the present paper 
we solved the linearized field equation for the stationary case, the same procedure may be 
applicable also for the time-dependent case. This will be studied in future work. 

The present analysis is based on the assumption that the linearized theory can be applied at 
 distances far from the source. The solution obtained by Mikhail et al.$^{21)}$ which violates $E=m$,
does not satisfy this assumption. It can also be shown that their solution does not satisfy the 
M\o ller condition (B), because the divergence  ${{\cal U_\alpha}^{0 \beta}}_{,\beta}$ is 
$O(1/r^{5/2})$ for their solution. 
\bigskip
\bigskip
\newpage
\centerline{\Large{\bf Acknowledgments}}

One of the authors (G. N.) would like to thank the Japanese Government for 
supporting him with a Monbusho Scholarship and also wishes to express his deep 
gratitude to all the members of Physics Department at Saitama University, 
 in particular,  Professor K. Kobayashi, Professor 
Y. Tanii, Professor K. Tanabe and Professor N. Yoshinaga. 

\bigskip
\bigskip
\centerline{\Large{\bf Appendix A}}
\centerline{\large Solution of an Inhomogeneous Laplace Equation}

Here we derive the 
general solution of an inhomogeneous Laplace equation of the form 
 $$
 \Delta {U(x,y,z)}={1 \over r^3}
 \sum_{N=1}^{\infty}{A_{\alpha_1 \alpha_2 \cdots {\alpha_ N}}} 
 n^{\alpha_1} n^{\alpha_2} \cdots n^{\alpha_N},
 \eqno{({\rm A} \cdot1)} \nonu
$$
%\ea 
where $A_{\alpha_1 \alpha_2 \cdots \alpha_N}$ is totally symmetric and traceless, and 
 $n^{\alpha}$ represents for the unit vector $x^{\alpha}/r$. 
We look for a solution in the form 
$$
U(x,y,z)={1 \over r}
 \sum_{N=0}^{\infty}{B_{\alpha_1 \alpha_2 \cdots \alpha_ N}} 
 n^{\alpha_1} n^{\alpha_2} \cdots n^{\alpha_N},
 \eqno{({\rm A} \cdot2)} \nonu
$$
%\ea 
where $B_{\alpha_1 \alpha_2 \cdots \alpha_N}$ is totally symmetric and traceless. 
Applying the Laplacian operator on (A$\cdot$2), and substituting the result in (A$\cdot$1)
 we obtain
$$
U(x,y,z)={B \over r} -{1 \over r}
 \sum_{N=1}^{\infty}{1 \over N(N+1)}{A_{\alpha_1 \alpha_2 \cdots \alpha_ N}} 
 n^{\alpha_1} n^{\alpha_2} \cdots n^{\alpha_N},
\eqno{({\rm A} \cdot3)} \nonu
$$
where $B$ is an arbitrary constant. 
Here we have used the following proposition:
$$
 \sum{A_{\alpha_1 \alpha_2 \cdots \alpha_N} n^{\alpha_1} n^{\alpha_2} \cdots
 n^{\alpha_N}}=0 \quad \Rightarrow A_{\alpha_1 \alpha_2 \cdots \alpha_N}=0,
\eqno{({\rm A} \cdot4)} \nonu
$$
 when $A_{\alpha_1 \alpha_2 \cdots \alpha_N}$ is totally symmetric and traceless. 
\bigskip
\bigskip

%%%%%%%%%%%%%%%%%%%%%%%%%%%%%%%%%%%%%%%%%%%%%%%%%%%%%%%%%%%%%%%%%%%%%%
%%%%%%%%%%%%%%%%%%%%%%%%%%%%%%%%%%%%%%%%%%%%%%%%%%%%%%%%%%%%%%%%%%%%%%

\newpage

\centerline{\Large{\bf References}}

\bigskip

\begin{enumerate}

\item[{1)}] A. Einstein, 
{\rm \rm S.\ B.\ Preuss.\  Akad.\ Wiss.\ }(1915), 778;  (1916), 1111.\

\item[{2)}] A. Einstein, 
{\rm S.\ B.\ Preuss.\  Akad.\ Wiss.\ } (1918), 448.\

\item[{3)}] F. Klein, 
{\rm G\"{o}tt.\ Nachr.\ Math.\  Phys.\ Klasse.\ } (1918), 394.\

\item[{4)}] R. M. Schoen and S. T. Yau,
{\rm Commun.\ Math.\  Phys.\ }{\bf65} (1979), 45; 
{\rm Phys.\ Rev.\  Lett.\ } {\bf48} (1982), 369.\

\item[{5)}] E. Witten, 
{\rm Comm.\ Math.\  Phys.\ }{\bf 80} (1981), 381.\

\item[{6)}] J. M.  Nester, 
{\rm Phys.\ Lett.\ }{\bf 83A} (1981), 241.\

\item[{7)}] C. M\o ller, 
{\rm Ann.\  of Phys.\ }{\bf 4}, (1958) 347.\

\item[{8)}] C. M\o ller, 
{\rm Mat.\ Fys.\ Medd.\ Dan.\ Vid.\ Selsk.\ }{\bf 35} (1966), no. 3.\\
The earlier version of M\o ller's requirements can be found in Ref. 10) cited below.

\item[{9)}] C. M\o ller, 
{\rm Mat.\ Fys.\ Medd.\ Dan.\ Vid.\ Selsk.\ }{\bf 31} (1959), no. 14.\

\item[{10)}] C. M\o ller, 
{\rm Ann.\ of Phys.\ }{\bf 12} (1961), 118.\

\item[{11)}] G. Lessner, 
{\rm  Gen.\ Rel.\ Grav.\ }{\bf 28} (1996), 527.\

\item[{12)}] C. M\o ller, 
{\rm  Mat.\ Fys.\ skr.\ Dan.\ Vid.\ Selsk.\ }{\bf 1} (1961), no. 10.\

\item[{13)}] C. M\o ller, 
{\rm Nucl.\ Phys.\ }{\bf 57} (1964), 330.\

\item[{14)}] R. Weitzenb$\ddot{o}$ck,
{\it Invariantentheorie \ }(Noordhoff, Groningen, 1923), p. 317.\

\item[{15)}] C. Pellegrini and J. Plebanski, 
{\rm Mat.\ Fys.\ skr.\ Dan.\ Vid.\ Selsk.\ }{\bf 2} (1963), no. 4.\

\item[{16)}] K. Hayashi and T. Nakano,
{\rm Prog.\ Theor.\ Phys.\ }{\bf 38} (1967), 491.\

\item[{17)}] C. M\o ller, 
{\rm Mat.\ Fys.\ Medd.\ Dan.\ Vid.\ Selsk.\ }{\bf 39} (1978), no. 13.\

\item[{18)}] K. Hayashi and T. Shirafuji, 
{\rm Phys.\ Rev.\ }{\bf D19} (1979), 3524; {\bf D24} (1981), 3312. \

\item[{19)}] A. Einstein, 
 {\rm S.\ B.\ Preuss.\ Akad.\ Wiss.\ } (1928), 217.\ 

\item[{20)}] J. Nitsch and F. W. Hehl, 
{\rm Phys.\ Lett.\ }{\bf 90B} (1980), 98.\

\item[{21)}] F. I. Mikhail, M. I. Wanas, A. Hindawi and E. I. Lashin, 
 {\rm Int.\ J.\  Theor.\ Phys.\ }{\bf 32} (1993), 1627.\

\item[{22)}] T. Shirafuji, G. G. L. Nashed and K. Hayashi, 
{\rm Prog.\ Theor.\ Phys.\ }{\bf 95} (1996), 665.\

\item[{23)}] T. Shirafuji, G. G. L. Nashed and Y. Kobayashi, 
{\rm Prog.\ Theor.\ Phys.\ }{\bf 96} (1996), 933.\

\item[{24)}] C. W. Misner, K. S. Thorne and J. A. Wheeler,
{\it Gravitation} (Freeman, San Francisco, 1973), p.\ 435.\

\end{enumerate}

%%%%%%%%%%%%%%%%%%%%%%%%%%%%%%%%%%%%%%%%%%%%%%%%%%%%%%%%%%%%%%%%%%%%%%%%%
%%%%%%%%%%%%%%%%%%%%%%%%%%%%%%%%%%%%%%%%%%%%%%%%%%%%%%%%%%%%%%%%%%%%%%%%%
%%%%%%%%%%%%%%%%%%%%%%%%%%%%%%%%%%%%%%%%%%%%%%%%%%%%%%%%%%%%%%%%%%%%%%%%%

\end{document}